 \documentclass{emulateapj}
 \usepackage{apjfonts}
\usepackage{graphicx}

\shortauthors{Winn et al.\ 2011}
\shorttitle{Transits of 55 Cnc~e}

\begin{document}

% ------------------------------------------------------------------------
% New commands
%
\def\ltsima{$\; \buildrel < \over \sim \;$}
\def\lsim{\lower.5ex\hbox{\ltsima}}
\def\gtsima{$\; \buildrel > \over \sim \;$}
\def\gsim{\lower.5ex\hbox{\gtsima}}
                                                                                          
% -------------------------------------------------------------------------
%

\bibliographystyle{apj}

\title{ A Super-Earth Transiting a Naked-Eye Star$^\star$ }

\author{
Joshua N.\ Winn\altaffilmark{1},
Jaymie M.\ Matthews\altaffilmark{2}, 
Rebekah I.\ Dawson\altaffilmark{3},
Daniel Fabrycky\altaffilmark{4,5},
Matthew J.\ Holman\altaffilmark{3},  \\
Thomas Kallinger\altaffilmark{2,6},
Rainer Kuschnig\altaffilmark{6},
Dimitar Sasselov\altaffilmark{3}, 
Diana Dragomir\altaffilmark{5},
David B.\ Guenther\altaffilmark{7},  \\
Anthony F.\ J.\ Moffat\altaffilmark{8},
Jason F.\ Rowe\altaffilmark{9},
Slavek Rucinski\altaffilmark{10},
Werner W.\ Weiss\altaffilmark{6}
}

 \journalinfo{Accepted version, July 6, 2011}
 \slugcomment{ApJ Letters, in press}

\altaffiltext{$\star$}{Based on data from the {\it MOST} satellite, a
  Canadian Space Agency mission, jointly operated by Dynacon Inc., the
  University of Toronto Institute for Aerospace Studies, and the
  University of British Columbia, with the assistance of the
  University of Vienna.}

\altaffiltext{1}{Department of Physics, and Kavli Institute for
  Astrophysics and Space Research, Massachusetts Institute of
  Technology, Cambridge, MA 02139}

\altaffiltext{2}{Department of Physics and Astronomy, University of
  British Columbia, 6224 Agricultural Road, Vancouver, BC V6T 1Z1,
  Canada}

\altaffiltext{3}{Harvard-Smithsonian Center for Astrophysics, 60
  Garden St., Cambridge, MA 02138}

\altaffiltext{4}{Hubble Fellow}

\altaffiltext{5}{UCO/Lick Observatory, University of California, Santa
  Cruz, CA 95064}

\altaffiltext{6}{University of Vienna, Institute for Astronomy,
  T\"urkenschanzstrasse 17, A-1180 Vienna, Austria}

\altaffiltext{7}{Department of Astronomy and Physics, Saint Mary's
  University, Halifax, NS B3H 3C3, Canada}

\altaffiltext{8}{D\'epartment de Physique, Universit\'e de Montr\'eal,
  C.P.\ 6128, Succ.\ Centre-Ville, Montr\'eal, QC H3C 3J7, Canada}

\altaffiltext{9}{NASA Ames Research Center, Moffett Field, CA 94035}

\altaffiltext{10}{Department of Astronomy \& Astrophysics, University
  of Toronto, 50 St.\ George Street, Toronto, ON M5S 3H4, Canada}

\begin{abstract}

  We have detected transits of the innermost planet ``e'' orbiting
  55~Cnc ($V=6.0$), based on two weeks of nearly continuous
  photometric monitoring with the {\it MOST} space telescope. The
  transits occur with the period (0.74~d) and phase that had been
  predicted by Dawson \& Fabrycky, and with the expected duration and
  depth for the crossing of a Sun-like star by a hot
  super-Earth. Assuming the star's mass and radius to be
  $0.963_{-0.029}^{+0.051}$~$M_\odot$ and $0.943\pm 0.010$~$R_\odot$,
  the planet's mass, radius, and mean density are $8.63\pm
  0.35$~$M_\oplus$, $2.00\pm 0.14$~$R_\oplus$, and
  $5.9_{-1.1}^{+1.5}$~g~cm$^{-3}$. The mean density is comparable to
  that of Earth, despite the greater mass and consequently greater
  compression of the interior of 55 Cnc~e. This suggests a rock-iron
  composition supplemented by a significant mass of water, gas, or
  other light elements. Outside of transits, we detected a sinusoidal
  signal resembling the expected signal due to the changing
  illuminated phase of the planet, but with a full range ($168\pm
  70$~ppm) too large to be reflected light or thermal emission. This
  signal has no straightforward interpretation and should be checked
  with further observations. The host star of 55~Cnc~e is brighter
  than that of any other known transiting planet, which will
  facilitate future investigations.

\end{abstract}


\section{Introduction}
\label{sec:introduction}

Precise Doppler observations have revealed five planets orbiting the
nearby G8~V star 55~Cnc (Butler et al.~1997, Marcy et al.~2002,
McArthur et al.~2004, Wisdom 2005, Fischer et al.~2008). Only a few
other stars are known to host as many planets: HD~10180 (Lovis et
al.~2011), Kepler-11 (Lissauer et al.~2011), and the Sun. Among the
other reasons why 55~Cnc has attracted attention are the 3:1 resonance
between two of its planets (Novak et al.~2003), the existence of an M
dwarf companion at a distance of 10$^3$~AU (Mugrauer et al.~2006), and
the unusually low mass and short period of its innermost planet,
designated ``e''.

McArthur et al.~(2004) reported a period and minimum mass for 55~Cnc~e
of 2.8~d and $14~M_\oplus$, respectively. Those parameters were
confirmed by Fischer et al.~(2008). More recently, Dawson \& Fabrycky
(2010) argued that 55~Cnc~e had been mischaracterized due to aliasing
in the radial-velocity data, and that the true period and minimum mass
are $0.74$~d and $8$~$M_\oplus$.

One implication of the shorter period would be an increased transit
probability, from 13\% to 25\%. The occurrence of transits enhances
the importance of an exoplanetary system, because transits can reveal
many details about the planet's dimensions, atmosphere, and orbit
(see, e.g., Winn 2010).

Fischer et al.~(2008) searched for transits in their 11-year
photometric record, ruling out transits for planets b ($P=14.7$~d) and
c (44.3~d). However, the time coverage was not complete enough to rule
out transits for planets f (260~d) and d (5200~d), and the precision
was insufficient to detect transits of the smallest planet e.

Here, we present space-based photometry of 55~Cnc revealing a transit
signal with the characteristics predicted by Dawson \& Fabrycky
(2010). Section 2 presents the data, and Section 3 presents the light
curve analysis, yielding estimates for the mass, radius, and density
of the planet. In Section 4 we place 55~Cnc~e in the context of the
small but growing population of super-Earths with measured masses and
radii.

While this manuscript was under review, we learned that Demory et
al.~(2011) detected a transit of 55~Cnc with the {\it Spitzer Space
  Telescope}. We refer the reader to that work for a complementary
analysis of the system properties.

\begin{figure*}[ht]
%\epsscale{1.0}
%\plotone{lc.eps}
 \begin{center}
 \leavevmode
 \hbox{
 \epsfxsize=6.5in
 \epsffile{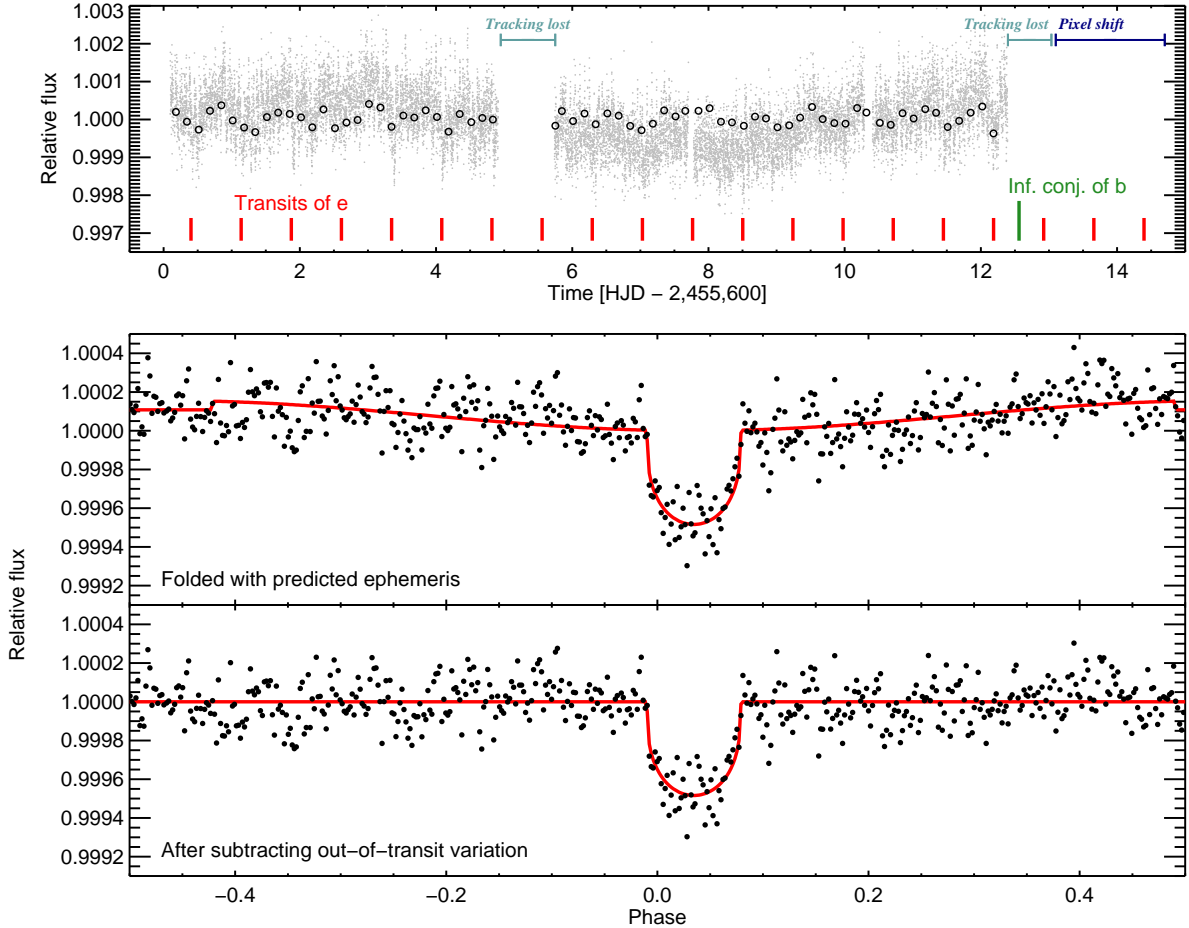}}
 \end{center}
 \vspace{-0.2in}
 \caption{ {\bf {\it MOST} photometry of 55~Cnc.}  {\it Upper.}---The
   time series, after decorrelation (small gray dots) and after
   further correction with the running averaged background method
   (large open circles, 0.25~d averages).  Vertical bars mark the
   predicted transit times of planet e, and the inferior conjunction
   of planet b (which was missed during a failure of fine
   tracking). {\it Middle.}---Phased light curve, folded with
   $P=0.736540$~d and $T_c$~[HJD]~$= 2,453,094.6924$ (Dawson \&
   Fabrycky 2010) and averaged into 2~min phase bins. The solid curve
   is the best-fitting model. {\it Bottom.}---Same, but with the
   best-fitting model of the out-of-transit variation has been
   subtracted from the data.
   \label{fig:lc}}
\end{figure*}

\section{Observations}
\label{sec:observations}

We observed 55~Cnc with {\it MOST} (Microvariability {\&} Oscillations
of STars), a Canadian microsatellite equipped with a 15~cm telescope
and CCD photometer, capable of short-cadence, long-duration
ultraprecise optical photometry of bright stars (Walker et al.\ 2003,
Matthews et al.\ 2004). {\it MOST} is in a Sun-synchronous polar orbit
820~km above the terminator with an orbital period of 101~min. Its
custom broadband filter covers the visible spectrum (350-700~nm).

We used the Direct Imaging mode, similar to conventional ground-based
CCD photometry. The observations were nearly continuous from
2011~February 07-22, except for a few interruptions when cosmic ray
hits during passages through the South Atlantic Anomaly resulted in
the loss of fine tracking. Individual exposures lasted 0.5~s but were
downloaded in stacks of 40 for the first 0.6~d, and stacks of 80 for
the remaining 14.4~d.

Aperture photometry was performed on the Direct Imaging subraster of
the Science CCD. Data affected by cosmic rays, image motion, or other
problems were identified and removed. To improve the homogeneity of
the data, we omitted data from the first 0.6~d (which had a different
effective exposure time) and the final 2.1~d (which suffered from a
tracking loss followed by a major shift in image registration). The
final time series has 18,373 data points and a time sampling of 43~s
outside of the interruptions.

Further processing was needed to remove the familiar periodic
artifacts in the time series due to scattered Earthshine. First, the
observed magnitude of 55~Cnc was fitted to a linear function of the
background level, $X$ position, and $Y$ position, and then this
function was subtracted from the observed magnitudes. The Fourier
spectrum still had significant peaks at the 14.26~c~d$^{-1}$ orbital
frequency of the satellite and its harmonics, as well as sidelobes at
$\pm$1~c~d$^{-1}$ away from those frequencies (arising from the
modulation of the stray light by the Earth's albedo pattern as viewed
by the satellite). For this reason, an additional correction was
performed with the ``running averaged background'' method of Rucinski
et al.~(2004). The data were divided into 5 time intervals, each
spanning approximately 32 {\it MOST} orbits (2.3~d). Within each
interval, the data were folded with the satellite's orbital period and
boxcar-smoothed, giving a reconstruction of the stray-light waveform
during that time interval. This waveform was then subtracted from the
observed magnitudes.

The upper panel of Figure~\ref{fig:lc} shows the final time series,
and the lower two panels show the data after folding with the Dawson
\& Fabrycky~(2010) ephemeris. A dip is observed at nearly zero phase,
where the transit signal would be expected. In addition, a gradual
rise in flux is observed away from zero phase, which is evident in the
middle panel, and which has been subtracted in the lower panel based
on the model described in \S~3.

\begin{figure}[ht]
\epsscale{1.2}
% \epsscale{1.0}
\plotone{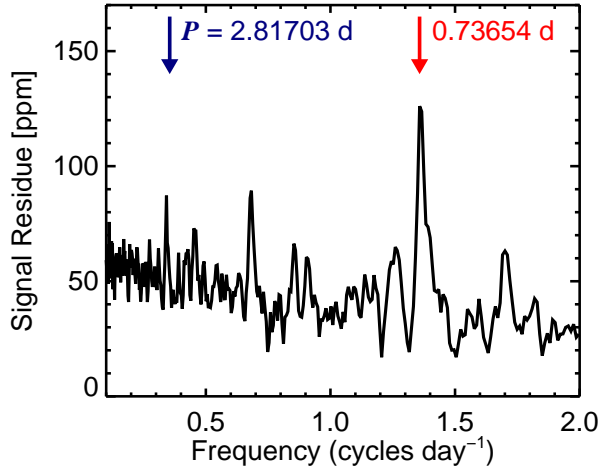}
\caption{{\bf Box-fitting Least Squares frequency spectrum of the {\it MOST} data.}
The spectrum was computed with the method of Kov{\'a}cs et al.~(2002). 
Positive peaks indicate detections of candidate transit signals with
durations consistent with a near-equatorial transit of 55~Cnc.
\label{fig:bls}}
\end{figure}

We emphasize that the signal shown in Figure~\ref{fig:lc} is not the
outcome of a period search: the data were phased with the {\it
  predicted} ephemeris. Nevertheless, when a period search is
performed the strongest signal is at 0.74~d, as shown in
Figure~\ref{fig:bls}. The signal has the predicted period, and the
observed epoch is bracketed by the two predicted epochs of Dawson \&
Fabrycky~(2010). It is 37~min later than the circular-orbit prediction
and 21~min earlier than the eccentric-orbit prediction. Furthermore
the depth and duration of the signal conform with expectations (see
\S~3). With close matches to four predicted parameters (period, phase,
depth, and duration) we consider the existence of transits to be
established.

\section{Analysis}
\label{sec:analysis}

\subsection{Light curve fitting}

A transit model was fitted to the light curve based on the formulas of
Mandel \& Agol (2002), and the Monte Carlo Markov Chain (MCMC) code of
Holman et al.\ (2006) and Winn et al.\ (2007). The orbit was assumed
to be circular, and the stellar limb-darkening law was assumed to be
quadratic. To model the out-of-transit variation seen in the middle
panel of Figure~\ref{fig:lc}, we added a term
\begin{equation}
F_{\rm pha} = \frac{\epsilon_{\rm pha}}{2}(1 - \cos2\pi\phi),
\end{equation}
where $\phi$ is the orbital phase relative to midtransit. For
completeness the model also included an occultation at $\phi=0.5$,
although occultations were not detected. The model parameters were the
planet-to-star radius ratio $R_p/R_\star$, star-to-orbit radius ratio
$R_\star/a$, orbital inclination $i$, time of midtransit $T_c$,
amplitude of the orbital phase modulation $\epsilon_{\rm pha}$,
occultation depth $\epsilon_{\rm occ}$, flux normalization (taken to
be the flux just outside of transit), and limb-darkening coefficients
$u_1$ and $u_2$.

Uniform priors were adopted for $R_p/R_\star$, $\cos i$, $T_c$,
$\epsilon_{\rm pha}$, $\epsilon_{\rm occ}$ and the flux
normalization. We used Gaussian priors on the stellar radius and mass,
$R_\star = 0.943\pm 0.010$~$R_\odot$ and $M_\star =
0.963_{-0.029}^{+0.051}~M_\odot$, which together act as a prior on
$R_\star/a$. The radius prior is based on the interferometrically
measured stellar radius (von Braun et al.~2011), and the mass prior is
based on the analysis of the stellar spectroscopic properties by
Takeda et al.\ (2007). Priors on the limb-darkening coefficients were
based on theoretical values $u_1 = 0.657$ and $u_2=0.115$, obtained by
integrating a Kurucz model with effective temperature 5327~K and $\log
g = 4.48$ over the {\it MOST} bandpass. The sum $u_1+u_2$ was subject
to a Gaussian prior with dispersion 0.1, and the difference $u_1-u_2$
(which has a negligible effect) was held fixed at the theoretical
value.

The likelihood was taken to be $\exp(-\chi^2/2)$ with the usual
sum-of-squares definition of $\chi^2$. The 1$\sigma$ uncertainty in
each data point was taken to be the root-mean-square (rms)
out-of-transit flux multiplied by a factor $\beta$ intended to take
into account the time-correlated noise. The factor $\beta$ is the
ratio between the standard deviation of residuals binned to 15~min,
and the standard deviation one would expect based on the unbinned data
assuming white noise (see, e.g., Pont et al.~2006, Carter \& Winn
2009). The rms and $\beta$ values were 101~ppm and 1.3,
respectively. Table~\ref{tbl:params} gives the results.

\subsection{Signal-injection tests}

To further investigate the effects of the correlated noise and
stray-light removal algorithms on the fitted transit parameters, we
injected and recovered fake transit signals. Beginning with the
aperture photometry, we subtracted the best-fitting transit model and
added a fake signal with a different period and transit time. The fake
signal had the same transit depth, duration (in phase units),
$\epsilon_{\rm pha}$ and $\epsilon_{\rm occ}$ as the best-fitting
model. Then, we processed the data just as was done with the
undoctored data. This was repeated for $10^3$ randomly chosen periods
within 50\% of the true period.

The recovered values of the transit and occultation depths had a
scatter of 47~ppm, in excess of the statistical error of 15~ppm, and
were systematically smaller by 1.9\% than the injected depths. The
fitted orbital phase modulations had a scatter of 68~ppm, in excess of
the statistical error of 15~ppm, and the amplitudes were 6.4\% smaller
than the injected values. Table~\ref{tbl:params} reports the values
after correcting for these biases and increased dispersions.

\section{ Discussion }
\label{sec:discussion}

\begin{figure*}[ht]
 \begin{center}
  \leavevmode
 \hbox{
  \epsfxsize=7.0in
  \epsffile{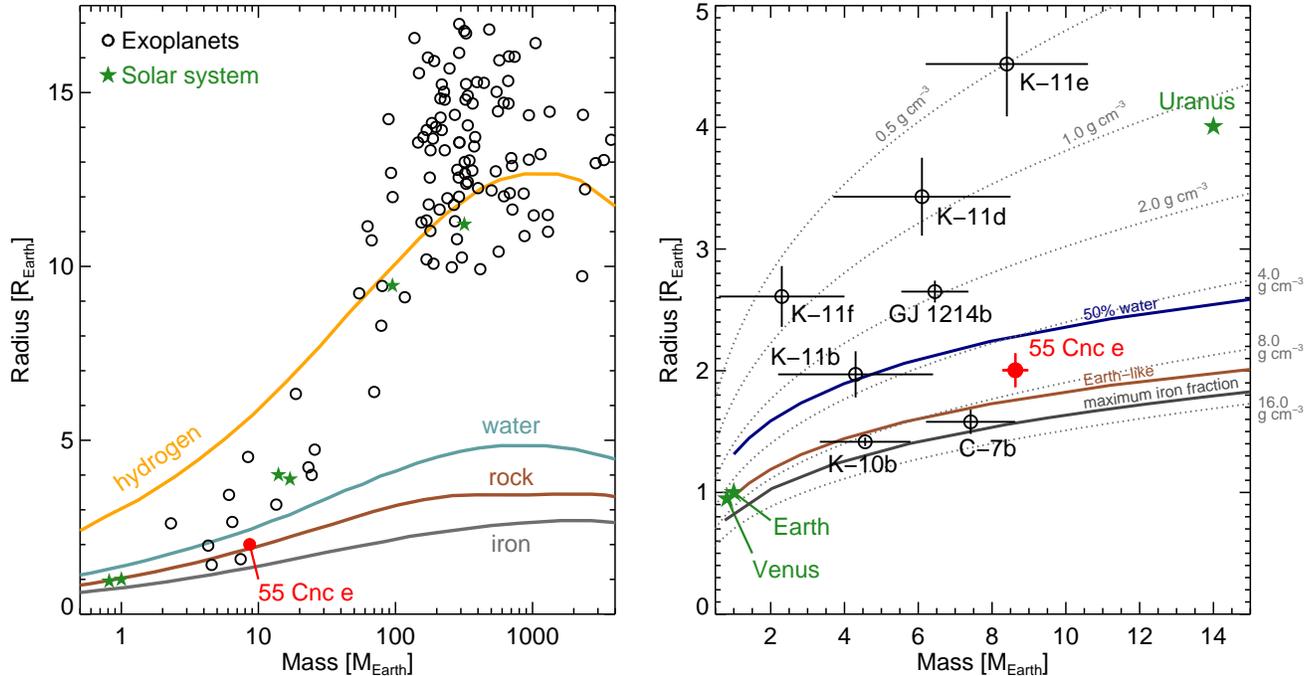}}
 \end{center}
 \vspace{-0.1 in}
%\epsscale{1.0}
%\plotone{dim.eps}
\caption{{\bf Masses and radii of transiting exoplanets.}
Open circles are previously known transiting planets. The filled
circle is 55~Cnc~e. The stars are Solar System planets,
for comparison. {\it Left.}---Broad view, with curves
showing mass-radius relations for pure hydrogen, water ice,
rock (MgSiO$_3$ perovskite)
and iron, from Figure~4 of Seager et al.~(2007).
{\it Right.}---Focus on super-Earths,
showing contours of constant mean density and a few
illustrative theoretical models: a ``water-world''
composition with 50\% water, 44\% silicate mantle
and 6\% iron core; a nominal ``Earth-like''
composition with terrestrial iron/silicon ratio and
no volatiles (Valencia et al.~2006, Li \& Sasselov, submitted);
and the maximum mantle stripping limit (maximum
iron fraction, minimum radius) computed by Marcus et al.~(2010).
Data were taken
from Lissauer et al.~(2011) for Kepler-11,
Batalha et al.~(2011) for Kepler-10b,
Charbonneau et al.~(2009) for GJ~1214b,
and Hatzes et al.~(2011) for Corot-7b.
We note the mass of Corot-7b is disputed (Pont et al.~2011).
\label{fig:mr}}
\vspace{0.3 in}
\end{figure*}

\subsection{Comparison to theoretical models}

Both the mass and radius of 55~Cnc~e are known to within 10\%,
providing a valuable example with which to test theoretical models of
super-Earth structure. To provide a broad view, the left panel of
Figure~\ref{fig:mr} shows the masses and radii of the transiting
exoplanets, along with theoretical curves taken from Seager et
al.~(2007) for ``mathematicians' planets'' composed of pure hydrogen,
water, rock (MgSiO$_3$ perovskite) and iron. The right panel focuses
on the super-Earths and shows the contours of constant mean density,
along with some theoretical curves based on more detailed models.

55~Cnc~e falls between the rock and water lines, suggesting it is
neither a gaseous planet, nor is it simply a scaled-up terrestrial
planet. Although the mean density of 55~Cnc~e is similar to that of
Earth, the greater compression of the interior of 55~Cnc~e implies
that it has a different composition. The {\it uncompressed} density of
55~Cnc~e would be smaller than that of Earth, implying that any rock
and iron must be accompanied by water, gas, or other light elements.

The right panel of Figure~\ref{fig:mr} also shows that the known
super-Earths span a factor of 20 in mean density, implying a
correspondingly large range of possibilities for composition and
internal structure. A striking contrast exists between 55~Cnc~e and
Kepler-11e, which have similar masses but densities differing by a
factor of 10.

\subsection{Atmosphere and orbital phase modulation}

Any atmosphere around 55~Cnc~e would be strongly heated, as the planet
is located less than 4~$R_\star$ from its host star. The planetary
temperature at the substellar point would be $T_\star\sqrt{R_\star/a}
\approx 2800$~K if the planet has a low albedo, its rotation is
synchronized with its orbit and the incoming heat is reradiated
locally. If instead the heat is redistributed evenly over the planet's
surface, the zero-albedo equilibrium temperature is
$T_\star\sqrt{R_\star/2a} \approx 1980$~K.

Atmospheres of transiting planets can be studied through occultations
and orbital phase variations (see, e.g., Knutson et al.~2007). Our
analysis did not reveal occultations ($\epsilon_{\rm occ} = 48\pm
52$~ppm), but did reveal a phase modulation ($\epsilon_{\rm pha} =
168\pm 70$~ppm). However, we cannot attribute the modulation to the
changing illuminated fraction of 55~Cnc~e, for two reasons. Firstly,
the occultation depth is smaller than the full range of the sinusoidal
modulation. Secondly, the amplitude of the modulation is too
large. Reflected starlight would cause a signal no larger than
$(R_p/a)^2 \approx 29$~ppm. The planet's thermal emission would
produce a signal $\approx$$(R_p/R_\star)^2(T_p/T_\star)^4 \approx
28$~ppm for bolometric observations, and only 5~ppm for observations
in the {\it MOST} bandpass, even for a 2800~K planet.

One possible explanation is that the star's planet-facing hemisphere
is fainter by a fraction $\epsilon_{\rm pha}$ than the other
hemisphere, due to star-planet interactions. The planet may induce a
patch of enhanced magnetic activity, as is the case for $\tau$~Boo~b
(Walker et al.~2008). In this case, though, the planet-induced
disturbance would need to be a traveling wave, because the stellar
rotation is not synchronized with the orbit. Fischer et al.\ (2008)
estimated the rotation period to be $42.7 \pm 2.5$~d, and Valenti \&
Fischer (2005) found the projected rotation speed to be $2.4\pm
0.5$~km~s$^{-1}$, much slower than the synchronous value of
65~km~s$^{-1}$.

Hence, the interpretation of the phase modulation is unclear. The
power spectral density of the photometric data also displays the
low-frequency envelope characteristic of stellar activity and
granulation, which complicates the interpretation of gradual
variations at the orbital period of 55 Cnc e. Confirming or refuting
this candidate orbital phase modulation is a priority for future work.

\subsection{Orbital coplanarity}
 
55 Cnc e is the innermost planet in a system of at least five
planets. If the orbits are coplanar and sufficiently close to
90$^\circ$ inclination, then multiple planets would transit. Transits
of b and c were ruled out by Fischer et al.~(2008).\footnote[11]{Our
  {\it MOST} observations might have led to firmer results for planet
  b, since it spanned a full orbit of that planet, but unfortunately
  no useful data were obtained during the transit window (see
  Fig.~\ref{fig:lc}). The {\it MOST} observation did not coincide with
  any transit windows for planets c-f.} However, the nondetections do
not lead to constraints on mutual inclinations. Given the measured
inclination for planet e of $90.0\pm 3.8$~deg, the other planets could
have orbits perfectly aligned with that of planet e and still fail to
transit.

McArthur et al.~(2004) reported an orbital inclination of $53^\circ
\pm 6.8^\circ$ for the outermost planet d, based on a preliminary
investigation of {\it Hubble Space Telescope} astrometry. This would
imply a strong misalignment between the orbits of d and e. However,
the authors noted that the astrometric dataset spanned only a limited
arc of the planet's orbit, and no final results have been
announced. Additional astrometric measurements and analysis are
warranted.

\begin{figure}[ht]
 \epsscale{1.1}
% \epsscale{1.0}
\plotone{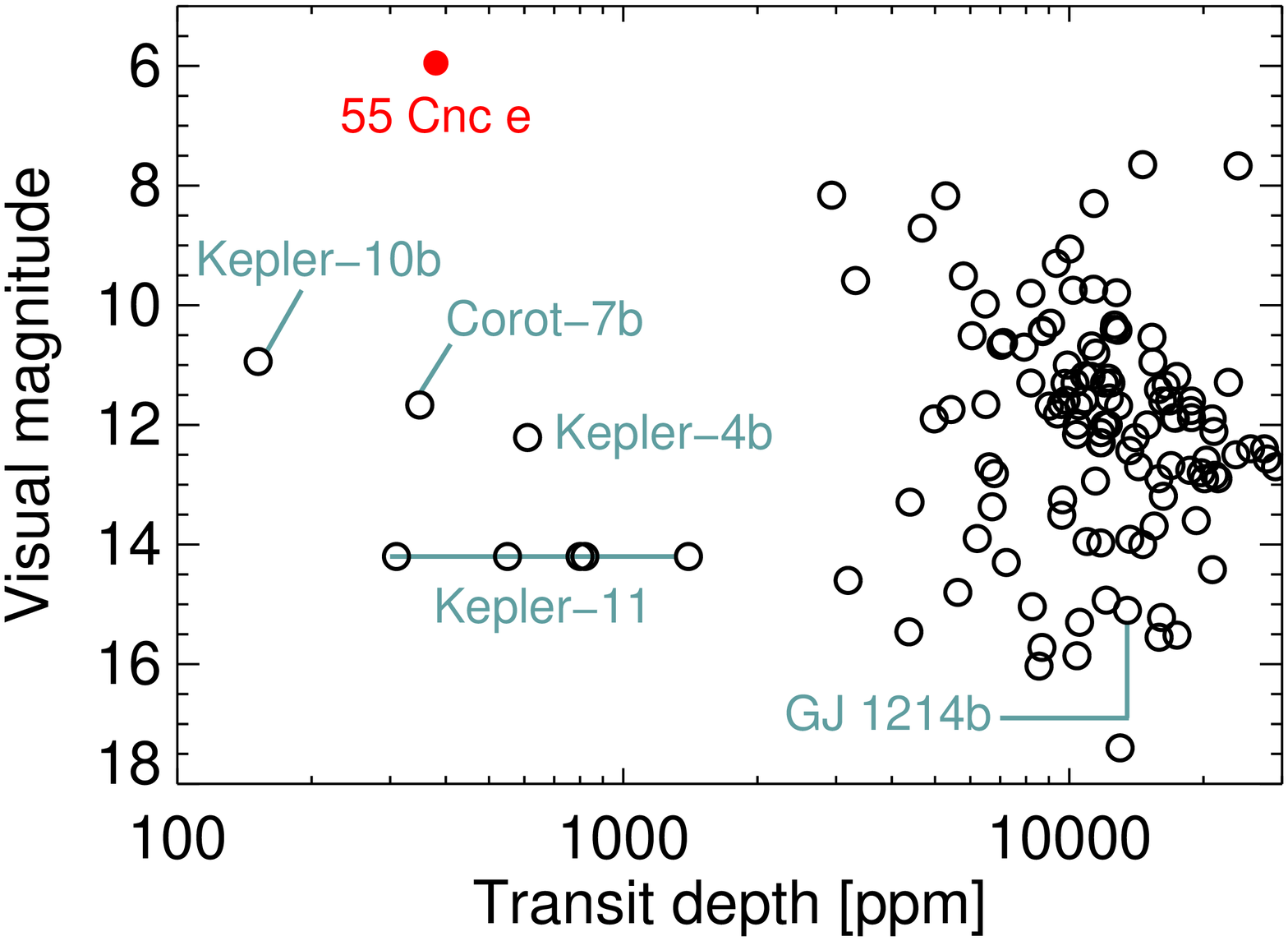}
\caption{{\bf Stellar brightness and transit depths.}
The $V$ band magnitudes and transit depths of
the transiting planets with known masses and radii.
Super-Earths ($M_p \lsim 10~M_\oplus$) are labeled.
\label{fig:obs}}
\end{figure}

\vspace{0.1in}
\subsection{Potential for follow-up observations}

Figure~\ref{fig:obs} shows the stellar brightness and transit depth
for each of the known transiting planets. 55~Cnc is a uniquely bright
host star, towering above the other super-Earth hosts and nearly 2~mag
brighter than any other transit star. However, Figure~\ref{fig:obs}
also shows that the transit depth for 55~Cnc~e is among the smallest
known. This combination of factors causes the follow-up landscape for
55~Cnc~e to differ from that of other planets.

The shallow depth will make certain follow-up observations challenging
despite the abundance of photons. To resolve the transit ingress and
egress, and thereby improve estimates of the planet's orbital
inclination and absolute dimensions, it will be necessary to improve
the signal-to-noise ratio in the phased light curve by observing more
transits or using a larger-aperture telescope. More data are also
needed to check on the candidate orbital phase modulation, and study
the atmosphere through occultation spectroscopy. Apart from
Kepler-10b, for which phase modulation was also tentatively detected
(Batalha et al.~2011), these effects have not yet been seen for
super-Earths.

Transit timing constraints on the system's architecture will not be
easily obtained, given the shallow transit and the small amplitudes of
the predicted signals. Even planet b, the nearest planet to e, is
expected to perturb e's transit epoch by less than 1~s over the course
of its 14~d period. The most readily detectable effect may be the
R\"omer delay due to planet d, which should cause a sinusoidal
variation in planet e's transit epoch with peak-to-trough amplitude of
$24$~s and period 5191~d.

On the other hand, follow-up observations of the star itself will
continue to be rewarding. Already the parallax and angular diameter of
the star have been measured, the stellar variability has been tracked
for 11 years (Fischer et al.~2008), and there is potential for the
detection of $p$-mode oscillations that would help define the stellar
properties (see, e.g., Gilliland et al.\ 2011, Nutzman et
al.~2011). The brightness of the star has already enabled the
discovery of 4 other planets in the system, and continued monitoring
has a greater potential to reveal additional bodies than is the case
for fainter stars.

Finally, there is some pleasure in being able to point to a naked-eye
star and know the mass and radius of one of its planets.

\acknowledgments We thank Laura McKnight and Andrew Howard for
thought-provoking conversations. Many people provided helpful feedback
on this work, including Simon Albrecht, Sarah Ballard, Rory Barnes,
Jacob Bean, Heather Knutson, David Latham, Barbara McArthur, Frederic
Pont, Daniel Rouan, Sara Seager, and the anonymous referee. We are
grateful to Kaspar von Braun for communicating the results of his
team's CHARA measurement prior to publication.

J.M., D.G., A.M., and S.R.\ thank NSERC~(Canada) for financial
support. T.K.\ is supported by a contract to the Canadian Space
Agency. R.K.\ and W.W.\ were supported by the Austrian Science
Fund. D.D.\ is supported by a FQRNT scholarship. R.D.\ is supported by
a National Science Foundation Graduate Research Fellowship, and
D.C.F.\ by NASA Hubble Fellowship HF-51272.01-A. M.H.\ and J.W.\ were
supported by NASA Origins award NNX09AB33G.\\

\begin{table*}
  \begin{center}
    \caption{System parameters for 55~\textup{Cnc~e}\label{tbl:params}}
    \smallskip 
    \begin{tabular}{l  r@{~$\pm$~}l}

      \tableline\tableline
      \noalign{\smallskip}
         Parameter & \multicolumn{2}{c}{Value} \\
      \noalign{\smallskip}
      \hline
      \noalign{\smallskip}

      Transit epoch~[HJD]                                & $2,455,607.05562 $ & $ 0.00087$ \\
      Transit depth, $(R_p/R_\star)^2$~[ppm]              & $380$ & $ 52$ \\
      Transit duration, first to fourth contact~[d]      & $0.0658 $ & $ 0.0013$ \\ 
      Transit ingress or egress duration~[d]             & $0.00134 $ & $ 0.00011$ \\
      \hline
      Planet-to-star radius ratio, $R_p/R_\star$          & $0.0195 $ & $ 0.0013$ \\  
      Transit impact parameter                           & $0.00$ & $ 0.24$ \\
      Orbital inclination, $i$~[deg]                     & $90.0$ & $ 3.8$  \\ 
      Fractional stellar radius, $R_\star/a$              & $0.2769 $ & $ 0.0043$ \\ 
      Fractional planetary radius, $R_p/a$               & $0.00539 $ & $ 0.00038$ \\ 
      Orbital distance, $a$~[AU]                         & $0.01583 $ & $ 0.00020$ \\
      \hline 
      Amplitude of orbital phase modulation, $\epsilon_{\rm pha}$~[ppm]  & $168$ & $70$ \\
      Occultation depth, $\epsilon_{\rm occ}$~[ppm]          & $48 $ & $ 52$ \\
      \hline 
      Planetary mass~[$M_\oplus$]                         & $8.63 $ & $ 0.35$ \\ 
      Planetary radius~[$R_\oplus$]                       & $2.00 $ & $ 0.14$ \\
      Planetary mean density~[g~cm$^{-3}$]                & $5.9$ & $_{1.1}^{1.5}$ \\
      Planetary surface gravity~[m~s$^{-2}$]              & $21.1$ & $_{2.7}^{3.5}$ \\

      \noalign{\smallskip}
      \tableline
      \noalign{\smallskip}
      \noalign{\smallskip}

    \end{tabular}
    
    \tablecomments{These parameters were determined by fitting the
      {\it MOST} light curve as described in the text, in combination
      with external constraints on the orbital period $P=0.7365400\pm
      0.0000030$~d and stellar reflex velocity $K_\star = 6.1\pm
      0.2$~m~s$^{-1}$ (Dawson \& Fabrycky 2010), stellar mass $M_\star
      = 0.963_{-0.029}^{+0.051}~M_\odot$ (Takeda et al.~2007), and
      stellar radius $R_\star = 0.943\pm 0.010$~$R_\odot$ (von Braun
      et al.~2011).  We further assumed the orbital eccentricity to be
      zero, and the limb-darkening law to be quadratic with
      coefficients $u_1$ and $u_2$ such that $u_1-u_2 = 0.542$ and
      $u_1+u_2=0.772\pm 0.100$.}
    
  \end{center}
  
\end{table*}

\end{document}